\documentclass[cameraready]{Interspeech}

\title{A Closer Look at Failure Modes in Temporal Understanding of \\Large Audio-Language Models}

\author[]{Apoorva}{Kulkarni}
\author[]{Kaousheik}{Jayakumar}
\author[]{Sreyan}{Ghosh}
\author[equalcontribution]{Sarah}{Wiegreffe}
\author[equalcontribution]{Dinesh}{Manocha}
\author[equalcontribution]{Ramani}{Duraiswami}

\address{
    University of Maryland, College Park, USA
}

\email{\{apoorvak, kajayaku, sreyang, sarahwie, dmanocha, ramanid\}@umd.edu}

\keywords{Large Audio Language Models, Interpretability, Temporal Reasoning, Audio Understanding}

\usepackage{comment}
\begin{document}

\maketitle
\begin{abstract}
Large Audio Language Models (LALMs) achieve strong performance on a variety of audio understanding tasks but continue to struggle with temporal reasoning, a fundamental capability central to human auditory perception. Understanding the causes of these failures remains challenging as existing benchmarks report performance gaps without probing underlying mechanisms. To address this, we introduce a benchmark with 1,657 questions across three foundational tasks designed specifically for mechanistic analysis. Examining model outputs across varying input settings (behavioral analysis) reveals that models often under-utilize audio when textual cues are available.  We also provide the \textbf{first causal mechanistic analysis} of temporal reasoning failures in LALMs. Comparing attention upweighting against scaling, we find that redistributing attention across audio tokens is more effective than increasing audio attention. Targeting task-relevant tokens yields further gains. These findings suggest that modality imbalance alone cannot explain failures. Attention scaling at bottleneck layers improves accuracy from 55.9\% to 59.1\% without fine-tuning, demonstrating a promising direction for future work.
\end{abstract}

\section{Introduction}
Large Audio Language Models (LALMs) have recently emerged as a key focus in multimodal AI, enabling a wide range of audio-centric tasks. Despite strong performance in identifying and describing acoustic events, models often struggle to localize events in time or reason about their temporal 
relationships~\cite{yao2025syncunveilingtemporalbias,Bhattacharya2025BenchmarkingACA}. These limitations reduce effectiveness in downstream tasks such as audio captioning with temporal grounding, sound-event detection, and diarization, where the order and duration of events determine meaning. Recent benchmarks such as MMAU~\cite{sakshi2024mmaumassivemultitaskaudio}, MMAR~\cite{ma2025mmarchallengingbenchmarkdeep} and MMAU-Pro~\cite{kumar2025mmauprochallengingcomprehensivebenchmark} confirm that temporal reasoning remains a challenge for state-of-the-art models. Despite this, relatively little work systematically investigates the mechanisms that cause temporal failures in LALMs. We present a controlled evaluation and mechanistic analysis of temporal reasoning in LALMs. Our contributions are:
\begin{itemize}[itemsep=0.2pt,topsep=0.2pt]
    \item Temporal reasoning benchmark with 1,657 questions across three foundational tasks: Earliest Onset, Latest Offset, and Longest Duration. These tasks target foundational capabilities for temporal reasoning. The narrow scope of the dataset is intentional and necessary for mechanistic analysis.
    \item Behavioral analysis shows that models under-utilize audio when textual cues are present. Attention patterns show text-dominant allocation across layers. These results are consistent with prior work.
    \item Application of causal attention interventions to temporal reasoning in LALMs. We compare attention upweighting, which increases total audio attention, against attention scaling, which redistributes attention across audio tokens. We find that attention scaling has greater impact in most configurations. Targeting task-relevant keyword tokens provides further benefit. This suggests that modality imbalance alone cannot explain failures: \textit{how} attention is distributed across audio tokens, not just \textit{how much} attention audio receives, is also a contributing factor.
    \item While our work is primarily diagnostic, we provide preliminary results on inference-time interventions. Attention scaling at identified bottleneck layers improves average accuracy (across models and tasks) from 55.9\% to 59.1\%, without requiring additional training data or model fine-tuning. This demonstrates that attention redistribution can enhance temporal reasoning and suggests a direction for future work.
\end{itemize}

\section{Related Work}
\noindent\textbf{LALM Benchmarks and Temporal Reasoning.}
Temporal reasoning has emerged as a key challenge for LALMs. Benchmarks such as MMAU~\cite{sakshi2024mmaumassivemultitaskaudio}, MMAU-Pro~\cite{kumar2025mmauprochallengingcomprehensivebenchmark}, and MMAR~\cite{ma2025mmarchallengingbenchmarkdeep} assess overall audio understanding across diverse tasks, with temporal reasoning as one component. Yao et al.~\cite{yao2025syncunveilingtemporalbias} systematically analyze how temporal reasoning varies with audio characteristics. Bhattacharya et al.~\cite{Bhattacharya2025BenchmarkingACA} examine performance and uncertainty of models on a synthetic temporal reasoning benchmark.

\noindent\textbf{Audio-Text Modality Imbalance.}
A prominent line of work has attributed LALM failures to modality imbalance. LALMs rely disproportionately on textual cues, sometimes overriding 
informative audio signals~\cite{wang-etal-2025-audio, rouditchenko2025omnir1reallyneedaudio}. This has motivated training-time methods that explicitly encourage audio contribution~\cite{he2025measuringaudiosimpactcorrectness}. However, these observations are derived from behavioral analysis, which does not establish causality.

\noindent\textbf{Mechanistic Interpretability for Multimodal Models.}
In vision-language models, Liu et al.~\cite{liu2024payingattentionimagetrainingfree} and Chen et al.~\cite{chen2025spatialreasoninghardvlms} demonstrate that model hallucinations and spatial reasoning failures are linked to attention mechanism failures. They propose training-free interventions to diagnose and mitigate these issues. Although fewer works focus specifically on audio, recent analyses of modality imbalance~\cite{wang2025payattentionaudiomitigating} examine attention mechanisms in LALMs. These studies motivate the use of mechanistic interpretability for LALMs.

\section{Dataset and Task Construction}
\label{sec:dataset}

\begin{figure*}[t]
    \centering
    \includegraphics[width=0.9\textwidth]{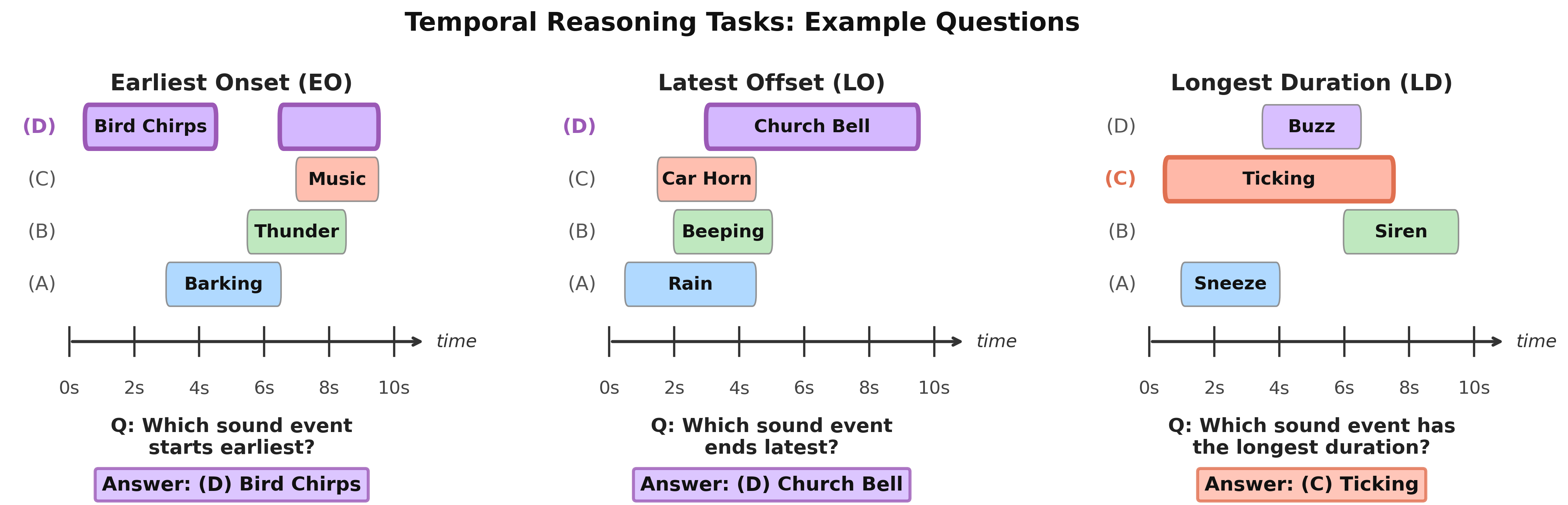}
    \caption{
    Example questions from the three temporal reasoning tasks with event timelines.
    Sound events may repeat or overlap, reflecting natural acoustic variation in real-world audio.
    Correct answers are highlighted.
    }
    \label{fig:task_examples}
\end{figure*}

We evaluate temporal reasoning using three controlled multiple-choice QA tasks derived from TACOS~\cite{primus2025tacostemporallyalignedaudiocaptions}, which provides temporally aligned audio segments with precise onset and offset annotations paired with textual descriptions. Each audio clip also includes a weak caption that provides general audio description. TACOS is sourced from Freesound, spanning 7 superclasses and 59 fine-grained categories, comprising real-world audio clips. Thus, the sound events may overlap or occur intermittently, increasing the difficulty of temporal reasoning.

Unlike broad benchmarks that test multiple skills, our goal is \textit{mechanistic diagnosis of a specific failure mode}: foundational temporal reasoning. By isolating minimal temporal capabilities, we can more precisely probe underlying model mechanisms.

\subsection{Tasks}
We design three tasks targeting temporal boundaries and duration. These tasks are prerequisites for higher-order reasoning.

\noindent\textbf{Earliest Onset (EO)} requires models to identify the sound event with the earliest start time among four options.

\noindent\textbf{Latest Offset (LO)} requires models to identify the sound event with the latest end time among four options.

\noindent\textbf{Longest Duration (LD)} requires models to identify the sound event with the longest duration among four options.

The dataset contains \textbf{1,657} questions: 528 EO, 499 LO, and 630 LD. \autoref{fig:task_examples} shows representative examples.

\subsection{Dataset Construction}

To construct the EO and LO tasks, we only retain instances where the correct event is separated from all others by at least one second. For the LD task, we only retain instances where the correct event is at least one second longer than all other events. Distractor options are generated in two stages: (1) include other events from the same clip; (2) if fewer than three are available, sample from different sound categories. All four options belong to distinct categories, and the correct answer is uniformly balanced across A/B/C/D.

\subsection{Validating Audio Contribution}

Following prior work~\cite{he2025measuringaudiosimpactcorrectness}, we evaluate whether our dataset requires strong audio-contribution. We perform silence ablation by replacing the audio input with silence for all tasks, to check the model’s reliance on textual priors. Across all models, accuracy drops to near chance, indicating that answers cannot be inferred from text alone. Because the weak caption is withheld here, this near-chance result is consistent with the higher caption-only accuracy in Section 4, where the caption supplies event cues the bare question lacks. The results are shown in \autoref{tab:silence_ablation}.

\begin{table}[h]
\centering
\resizebox{\columnwidth}{!}{
\begin{tabular}{lccc}
\toprule
\textbf{Model} & \textbf{EO (\%)} & \textbf{LO (\%)} & \textbf{LD (\%)} \\
\midrule
Qwen2-Audio-7B-Instruct & 30.87 & 28.06 & 32.54 \\
Kimi-Audio-7B-Instruct & 21.97 & 24.05 & 25.87 \\
Audio-Flamingo-3 & 25.57 & 20.64 & 28.89 \\
DeSTA2.5-Audio-Llama-3.1-8B & 24.43 & 25.05 & 22.70 \\
\midrule
Random baseline & 25.00 & 25.00 & 25.00 \\
\bottomrule
\end{tabular}
}
\caption{Audio contribution verification via silence ablation. All models achieve near-chance performance, confirming that correct answers cannot be inferred from text alone.}
\label{tab:silence_ablation}
\end{table}

\section{Behavioral Analysis}
\label{sec:behavioral}
Behavioral analysis is an interpretability approach that seeks to understand a model by systematically prompting it and observing its outputs under controlled conditions. We adopt this approach to examine how LALMs utilize audio versus textual information for temporal reasoning.

We evaluate four state-of-the-art open-source LALMs: Qwen2-Audio-7B-Instruct~\cite{chu2024qwen2audiotechnicalreport}, Kimi-Audio-7B-Instruct~\cite{kimiteam2025kimiaudiotechnicalreport}, Audio-Flamingo-3~\cite{goel2025audioflamingo3advancing}, and DeSTA2.5-Audio-Llama-3.1-8B~\cite{lu2025desta25audiogeneralpurposelargeaudio}. Following prior work~\cite{NEURIPS2024_89cc5e61}, we design three input formats to isolate and compare the contribution of each modality: (1) \textbf{Audio-only (AQA)}: audio input with question but without caption, (2) \textbf{Caption-only (CQA)}: weak caption text with question but without audio, (3) \textbf{Audio-Caption (ACQA)}: both audio and caption together with question. Results are shown in \autoref{tab:modality_comparison}. The analysis reveals two main findings.

\begin{table*}[t]
\centering
\small
\begin{tabular}{lcccccccccc}
\toprule
& \multicolumn{3}{c}{\textbf{Earliest Onset (EO)}} & \multicolumn{3}{c}{\textbf{Latest Offset (LO)}} & \multicolumn{3}{c}{\textbf{Longest Duration (LD)}} \\
\cmidrule(lr){2-4} \cmidrule(lr){5-7} \cmidrule(lr){8-10}
\textbf{Model} & \textbf{AQA} & \textbf{CQA} & \textbf{ACQA} & \textbf{AQA} & \textbf{CQA} & \textbf{ACQA} & \textbf{AQA} & \textbf{CQA} & \textbf{ACQA} \\
\midrule
Qwen2-Audio-7B-Instruct & 30.87 & 63.64 & 63.64 & 28.06 & 46.49 & 46.49 & 32.54 & 58.89 & 58.89 \\
Kimi-Audio-7B-Instruct & 57.95 & 61.36 & 68.75 & 60.32 & 56.11 & 62.73 & 59.37 & 55.40 & 67.30 \\
Audio-Flamingo-3 & 60.42 & 71.59 & 67.23 & 56.71 & 62.32 & 63.33 & 58.10 & 66.67 & 66.35 \\
DeSTA2.5-Audio-Llama-3.1-8B & 50.38 & 68.94 & 62.69 & 51.30 & 59.32 & 59.12 & 54.92 & 66.03 & 66.83 \\
\bottomrule
\end{tabular}
\caption{Performance (\%) across three modality settings on all tasks. AQA: audio-only input; CQA: caption-only input; ACQA: audio+caption input. CQA consistently outperforms AQA for most tasks and models, indicating reliance on textual cues.}
\label{tab:modality_comparison}
\end{table*}

\noindent\textbf{Temporal reasoning is challenging for LALMs.} 
Although models like Audio-Flamingo-3 and Kimi-Audio-7B-Instruct perform well on broad benchmarks such as MMAU, they achieve significantly lower accuracy on our temporal reasoning benchmark.

\noindent\textbf{Models under-utilize audio when text is available.} 
Despite our benchmark explicitly requiring audio contribution, models rely heavily on textual cues when available. Across most models and tasks, CQA outperforms AQA, often substantially. With the exception of Kimi-Audio, ACQA provides minimal benefit over CQA and in some cases reduces performance. This suggests that most models fail to effectively integrate audio information to refine temporal reasoning.

\begin{figure}[t]
\centering
\includegraphics[width=\columnwidth]{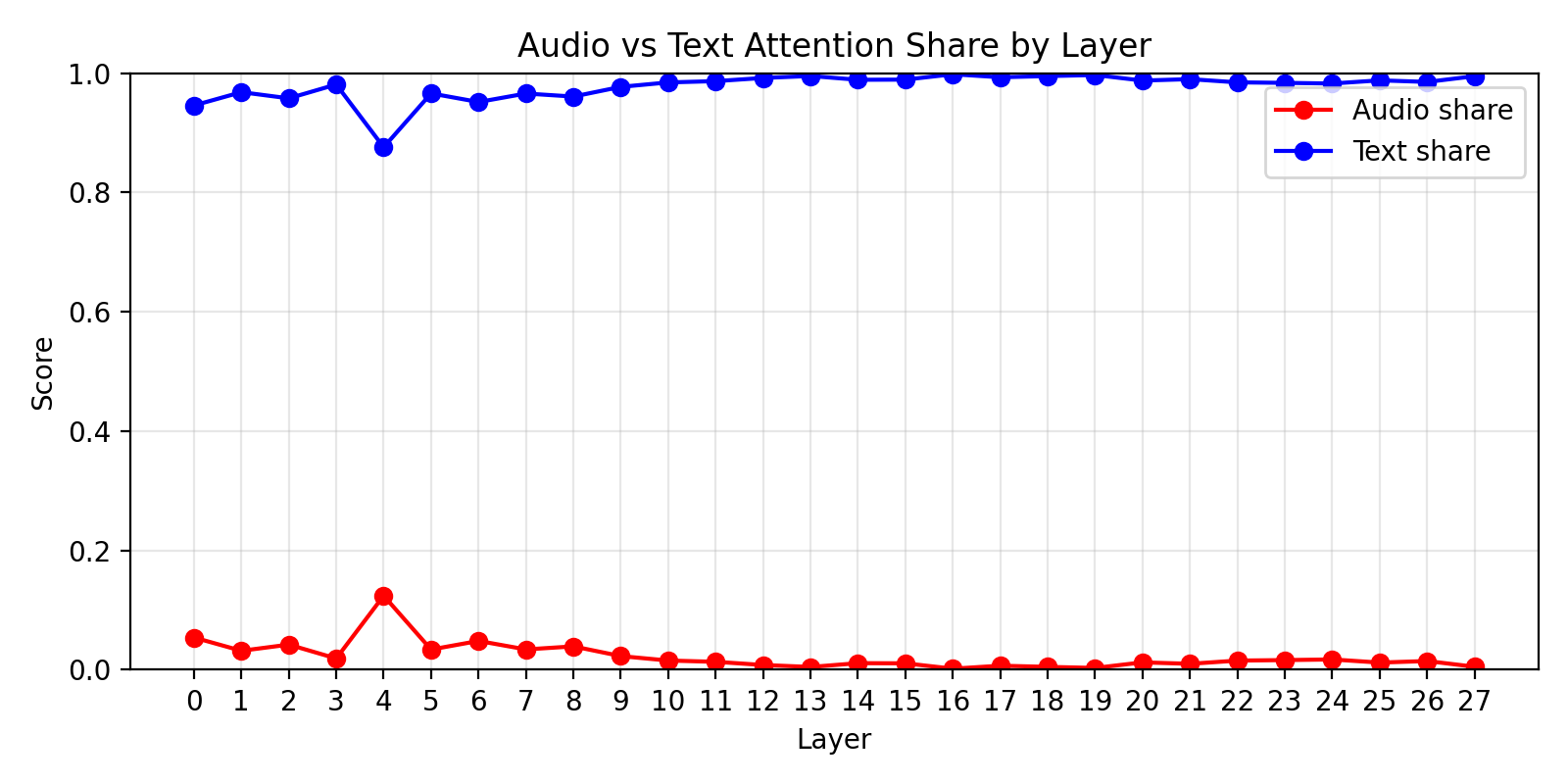}
\caption{Layer-wise attention distribution between audio and text modalities for Audio-Flamingo-3 for Earliest Onset (EO) task. Other models and tasks also exhibit similar text-dominant attention patterns.}
\label{fig:attention_patterns}
\end{figure}

To further examine modality utilization, we analyze attention patterns. For each layer, we compute the proportion of attention allocated to audio versus text tokens from the final input token, averaged across all attention heads and instances. \autoref{fig:attention_patterns} shows layer-wise attention distributions for Audio-Flamingo-3 for Earliest Onset (EO) task, which exhibits text-dominant attention patterns across most layers. Other models and tasks also exhibit similar text-dominant attention patterns. This is consistent with prior observations of modality imbalance in LALMs.

Both behavioral analysis and attention patterns provide evidence of text-dominant attention in LALMs. However, these observations are correlational and insufficient to establish causality. To move beyond correlation, we apply mechanistic interpretability techniques in the following section. By \textit{causally intervening} on attention mechanisms, we can test whether correcting audio-text imbalance resolves temporal reasoning failures, or whether other attention dynamics are at play.

\section{Mechanistic Analysis}
\label{sec:mechanistic}
For mechanistic analysis, we only utilize Audio-Flamingo-3 and DeSTA-2.5-Audio. These are the only state-of-the-art LALMs with fully open-source weights, training code, and training data. This enables reproducible mechanistic analysis and rules out data-driven confounds. We compare two training-free attention interventions, applied uniformly across all attention heads and layers:

\noindent\textbf{Attention Upweighting} increases total attention mass allocated to audio tokens. Inspired from prior work on vision, ~\cite{liu2024payingattentionimagetrainingfree, 
wang2025payattentionaudiomitigating} we amplify the pre-softmax attention logits from the final prompt token to all audio tokens:
\begin{equation}
\tilde{A}^{(\ell,h)}_{n,j} =
\begin{cases}
A^{(\ell,h)}_{n,j} + \alpha \left|A^{(\ell,h)}_{n,j}\right|, & j \in \mathcal{I}_{\text{audio}},\ \forall\, \ell,h\\
A^{(\ell,h)}_{n,j}, & \text{otherwise}
\end{cases}
\end{equation}
where $\ell$ indexes transformer layers, $h$ indexes attention heads, $n$ indexes the query position corresponding to the final prompt token, $j$ indexes key positions, $\mathcal{I}_{\text{audio}}$ denotes the set of audio-token indices, and $\alpha$ controls the strength of upweighting. 

\noindent\textbf{Attention Scaling (ScalingVis)} redistributes attention across audio tokens by multiplicatively scaling logits~\cite{chen2025spatialreasoninghardvlms}. Specifically, it targets attention from the final input token to all audio tokens, scaling those logits by a coefficient $\alpha$ to sharpen ($\alpha > 1$) or smooth ($\alpha < 1$) the attention 
distribution:
\begin{equation}
\tilde{A}^{(\ell,h)}_{n,j} =
\begin{cases}
\alpha \, A^{(\ell,h)}_{n,j}, & j \in \mathcal{I}_{\text{audio}} \\
A^{(\ell,h)}_{n,j}, & \text{otherwise}
\end{cases}
\end{equation}
ScalingVis is motivated by the intuition that if an attention pattern is broadly correct but lacks precision, sharpening may improve performance, whereas smoothing may be beneficial if the attention pattern is inherently misaligned.

\subsection{Results}
We evaluate interventions on samples where the model initially predicts incorrectly. We report \textbf{\textit{fix rate}}: the percentage of these incorrect predictions that are corrected after intervention. We apply interventions from three possible token position settings to all audio tokens: (1) \textbf{Last}: the final prompt token only, following prior work, (2) \textbf{Keyword}: task-relevant keyword tokens only (e.g., ``earliest'', ``latest'', ``longest''), and (3) \textbf{Kwd+Last}: both keyword and final prompt tokens. \autoref{tab:intervention_results} reports fix rates across both models and all tasks. Three findings emerge:

\begin{table}[t]
\centering
\small
\setlength{\tabcolsep}{2pt}
\begin{tabular}{ll|cc|cc}
\toprule
& & \multicolumn{2}{c|}{\textbf{Upweight}} & \multicolumn{2}{c}{\textbf{Scale}} \\
\textbf{Task} & \textbf{Tokens} & $\alpha$=0.1 & $\alpha$=0.5 & $\alpha$=0.2 & $\alpha$=2.0 \\
\midrule
\multicolumn{6}{c}{\textit{Audio-Flamingo-3}} \\
\midrule
EO & Last & 1.9 & 11.3 & 12.3 & 16.0 \\
EO & Keyword & 4.2 & 12.3 & 7.5 & 10.4 \\
EO & Kwd+Last & 6.1 & 11.8 & 10.8 & \textbf{18.4} \\
\midrule
LO & Last & 4.6 & 14.2 & 14.2 & 20.6 \\
LO & Keyword & 5.0 & 14.7 & 8.7 & 17.0 \\
LO & Kwd+Last & 3.7 & 14.7 & 12.8 & \textbf{24.3} \\
\midrule
LD & Last & 4.2 & 18.3 & 12.2 & \textbf{22.5} \\
LD & Keyword & 4.6 & 13.0 & 6.9 & 6.1 \\
LD & Kwd+Last & 5.3 & 19.8 & 15.6 & 18.7 \\
\midrule
\multicolumn{6}{c}{\textit{DeSTA-2.5-Audio}} \\
\midrule
EO & Last & 1.9 & 7.3 & 20.4 & 7.3 \\
EO & Keyword & 3.5 & 5.0 & 9.6 & 5.0 \\
EO & Kwd+Last & 3.5 & 8.1 & \textbf{21.2} & 6.5 \\
\midrule
LO & Last & 2.9 & 10.5 & 16.3 & 14.2 \\
LO & Keyword & 3.3 & 6.3 & 8.8 & 7.1 \\
LO & Kwd+Last & 4.6 & 12.1 & \textbf{20.1} & 13.4 \\
\midrule
LD & Last & 1.5 & 8.1 & \textbf{18.9} & 9.3 \\
LD & Keyword & 1.1 & 2.2 & 7.8 & 3.0 \\
LD & Kwd+Last & 1.9 & 8.1 & \textbf{18.9} & 9.3 \\
\bottomrule
\end{tabular}
\caption{Fix rate (\%) for attention interventions applied across all layers and heads on incorrectly predicted samples.}
\label{tab:intervention_results}
\end{table}

\noindent\textbf{Scaling outperforms upweighting.} Across both models, redistributing attention via scaling yields higher fix rates than increasing audio attention via upweighting. For Audio-Flamingo-3, scaling with $\alpha=2.0$ (sharpening) achieves 20.5\% average fix rate compared 
to 15.8\% for the best upweighting configuration. For DeSTA-2.5-Audio, scaling with $\alpha=0.2$ (smoothing) achieves 20.1\% average fix rate compared to 10.1\% for upweighting. This suggests that the imbalance hypothesis alone is insufficient: \textit{how} attention is distributed across 
audio tokens matters more than \textit{how much} total attention audio receives.

\noindent\textbf{Combining keyword and final tokens is most effective.} 
Applying interventions from both task-relevant keyword tokens and the final prompt token (Kwd+Last) yields the highest fix rates for both models. Keyword-only interventions are less effective than final-token-only, but combining both provides complementary benefit.

\noindent\textbf{Optimal intervention is architecture-dependent.} 
Audio-Flamingo-3 benefits from sharpening ($\alpha=2.0$), suggesting attention that is correctly directed but imprecise. DeSTA-2.5-Audio benefits from smoothing ($\alpha=0.2$), suggesting misaligned attention that requires redistribution.

\subsection{Preliminary Inference-Time Intervention}
Our mechanistic analysis identifies that attention scaling can correct a portion of temporal reasoning errors. We investigate whether targeted interventions can serve as a practical inference-time mitigation strategy.

\noindent\textbf{All-layer intervention degrades performance.} 
Applying scaling across all layers and all attention heads simultaneously causes significant disruption, resulting in accuracy drops across all configurations (\autoref{tab:all_layer_intervention}). We hypothesize that intervening across all layers is too broad, causing degredation to correctly-predicted data points.

\begin{table}[t]
\centering
\small
\setlength{\tabcolsep}{3pt}
\begin{tabular}{lcccc}
\toprule
& \multicolumn{2}{c}{\textbf{AF3}} & \multicolumn{2}{c}{\textbf{DeSTA}} \\
\cmidrule(lr){2-3} \cmidrule(lr){4-5}
\textbf{Task} & Baseline & All-Layer & Baseline & All-Layer \\
\midrule
EO & 59.84 & 51.13 & 50.75 & 51.70 \\
LO & 56.31 & 48.09 & 52.10 & 51.10 \\
LD & 58.41 & 47.77 & 57.14 & 52.22 \\
\bottomrule
\end{tabular}
\caption{Accuracy (\%) with all-layer, all-head scaling intervention. All-layer intervention degrades performance compared to baseline across both models and all tasks.}
\label{tab:all_layer_intervention}
\end{table}

\begin{figure}[t]
\centering
\includegraphics[width=0.9\columnwidth]{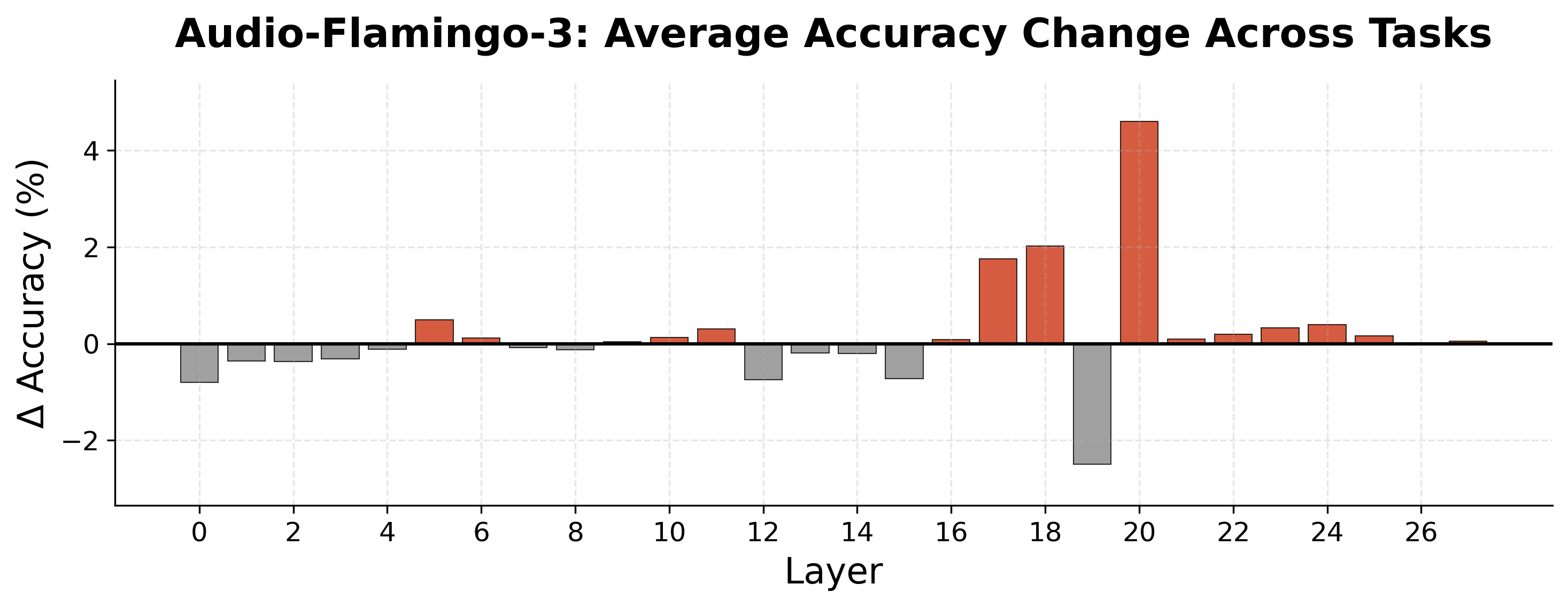}
\vspace{0.3em}
\includegraphics[width=0.9\columnwidth]{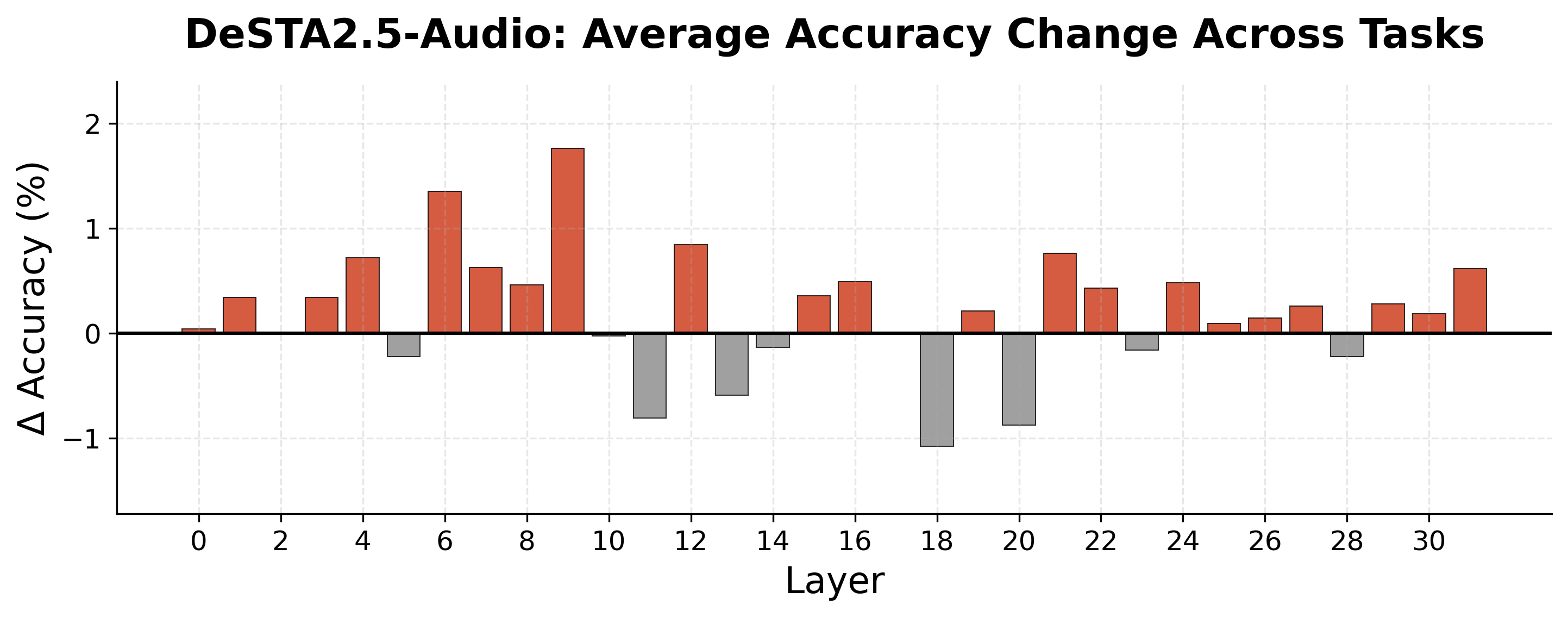}
\caption{Layer-wise scaling effect on accuracy. Audio-Flamingo-3 (top figure) exhibits peak improvement at Layer 20. DeSTA-2.5-Audio (bottom figure) shows peak improvement at Layer 9.}
\label{fig:layerwise_scaling}
\end{figure}

\noindent\textbf{Layer-targeted intervention improves performance.} 
We apply the best scaling intervention for each model at a single layer. \autoref{fig:layerwise_scaling} shows layer-wise accuracy changes under scaling for both models. Audio-Flamingo-3 exhibits a clear localized improvement at Layer 20 under sharpening ($\alpha=2.0$). DeSTA-2.5-Audio shows more distributed effects, with peak improvement at Layer 9 under smoothing ($\alpha=0.2$). Averaging across both models, layer-targeted scaling yields a 3.2\% improvement in temporal reasoning accuracy. These gains are modest but notable given that they require no additional training data, fine-tuning, or architectural modifications. This suggests that inference-time attention redistribution may be a useful direction for improving temporal reasoning in scenarios where training compute or data are limited.

\section{Limitations and Future Work}
Our attention-level interventions cannot rule out the impact of alternative mechanisms such as weak audio encoder representations. The fix rates achieved indicate that attention distribution is one contributing factor among others. However, our findings do rule out one prominent hypothesis: prior work has emphasized audio-text modality imbalance as a key failure mode, yet we show that simply increasing audio attention is less effective than redistributing it. We do not claim to identify the complete causal pathway of temporal reasoning failures, but provide diagnostic evidence shifting focus toward finer-grained attention dynamics. Future work could develop training-time interventions informed by these findings and extend the analysis to additional architectures and more complex temporal reasoning tasks.

\section{Conclusion}
This work investigates temporal reasoning failures in LALMs through a controlled benchmark with 1,657 questions across three foundational tasks. Behavioral analysis confirms models under-utilize audio when textual cues are available. We provide the first causal attention interventions for temporal reasoning in LALMs, adapting ScalingVis from vision-language models. Our key finding: redistributing attention via scaling outperforms simply increasing audio attention, and targeting task-relevant keyword tokens provides additional benefit. Preliminary layer-targeted interventions yield modest accuracy gains without additional training or data.

\section{Generative AI Use Disclosure}
\label{app:ai_use}
We utilized AI assistants to help clarify explanations, suggest concise phrasing, and organize text for readability. These tools were used exclusively for linguistic support and were not used to generate scientific results or formulate claims. 

\bibliographystyle{IEEEtran}
\bibliography{mybib}

\end{document}